\documentclass[12pt]{article}
\usepackage{color}
\usepackage{graphicx}

\begin{document}

\title{ ANTHROPIC PRINCIPLE IN COSMOLOGY  \\[1cm]
}

 \author{{\bf Brandon  Carter} \\
 {LuTh, Observatoire de Paris-Meudon }
  }
\date{\it Contribution to Colloquium ``Cosmology: Facts and problems''\\
 (Coll\`ege de France, June 2004).}
\maketitle

\vskip 1.2 cm

\noindent{\bf Abstract.}
\medskip
\parindent=2 cm

A brief explanation of the meaning of the anthropic principle -- as a 
prescription for the attribution of a priori probability weighting -- 
is illustrated by various cosmological and local applications, in which 
the relevant conclusions are contrasted with those that could be obtained 
from (less plausible) alternative prescriptions such as the vaguer and 
less restrictive ubiquity principle, or the more sterile and restrictive 
autocentric principle.

\vfill\eject

\noindent{\bf Introduction.}
\medskip
\parindent=2 cm

Having been asked to contribute a discussion of the anthropic principle 
for a colloquium on cosmology, I would start by recalling that although 
its original formulation \cite{Carter74} was motivated by a problem of 
cosmology (Dirac's) and although many of its most interesting subsequent 
applications (such as the recent evaluation~\cite{GarVil03} of the dark 
energy density in the universe)  have also been concerned with large scale 
global effects, the principle for which I introduced the term ``anthropic'' 
is not intrinsically cosmological, but just as relevant on small local 
scales as at a global level. In retrospect I am not sure that my choice of 
terminology was the most appropriate, but as it has now been widely 
adopted~\cite{Bostrom} it is too late to change. Indeed the term 
``anthropic principle'' has become so popular that it has been borrowed 
to describe ideas (e.g. that the universe was teleologically designed for 
our kind of life, which is what I would call a ``finality principle'') that 
are quite different from, and even contradictory with, what I intended. 
This presentation will not attempt to deal with the confusion that has 
arisen from such dissident interpretations, but will be concerned only 
with developments of my originally intended meaning, which I shall attempt 
to explain in the next section.

\bigskip

\noindent{\bf Meaning of the Anthropic principle.}
\medskip
\parindent=2 cm

Whenever one wishes to draw general conclusions from observations restricted
to a small sample it is essential to know whether the sample should be
considered to be biassed, and if so how. The anthropic principle provides
guidelines for taking account of the kind of biass that arises from the
observer's own particular situation in the world.

Although frequently relevant to purely local applications, the anthropic
principle was originally formulated in a cosmological context as a 
reasonable compromise two successively fashionable extremes. The first of 
these was what might be described as the autocentric principle, which 
underlay the pre Copernican dogma to the effect that as terrestrial 
observers we occupy a privileged position at the center of the universe. 
The opposite extreme was the more recent precept describable as the 
cosmological ubiquity principle, but commonly referred to just as the 
cosmological principle, which would have it that the Universe is much the 
same everywhere, having no priviledged center, and that our own 
neighbourhood can be considered as a typical random sample. 

To put it more formally, in conventional Bayesian terminology, the a priori
probability distribution for our own situation was supposed, according to 
the autocentric principle, to have been restricted to the region  where we
actually find ourselves, whereas according to the ubiquity principle it was
supposed to have been uniformly extended over the whole of space time. Thus
according to the autocentric principle we could infer nothing at all about 
the rest of the universe from our local observations, whereas according to 
the ubiquity principle we could immediately infer that the rest of the 
universe was fairly represented by what we observe here and now.

As a reasonable compromise between these unsatisfactory over simplistic 
extremes, the anthropic principle would have it that -- within the context
of whatever theoretical model may be under consideration -- the a priori 
probability distribution for our own situation should be prescribed by 
an anthropic weighting, meaning that it should be uniformly distributed, 
not over space time (as the ubiquity
principle would require), but over all observers sufficiently comparable 
with ourselves to be qualifiable as anthropic. 

Of course if the qualification ``anthropic'' were interpreted so narrowly 
as to include only members of our own human species, then the cosmological
implications of the anthropic principle would reduce to those of the
scientifically sterile autocentric principle, but it is intended that the 
term ``anthropic'' should also include extraterrestrial beings with 
comparable intellectual capabilities. Thus, unlike the autocentric principle 
but like the ubiquity principle, the anthropic principle has non trivial 
implications that can be subjected to empirical verification. The  prototype
 example was provided by the famous debate~\cite{Dicke61} between Dirac and 
Dicke about whether the strength of gravitation should decrease in 
proportion to the expansion of the universe: subsequent work has shown 
rather conclusively that Dirac's prediction (that it would), which was 
implicitly based on the cosmological ubiquity principle, must be rejected 
in favour of Dicke's prediction (that it would not), which was implicitly 
based on the anthropic principle. (This debate illustrates a common source 
of misunderstanding in this area, which is that relevant but questionable 
principles tend to be taken for granted tacitly, and even subconsciously, 
rather than being made explicit.)

If it were necessary to be more precise, one would need some kind of 
{\it  microanthropic} principle specifying the notion of anthropic 
weighting in greater detail, dealing with questions such as whether it 
should be proportional to the longevity and erudition of the individuals 
under consideration. (For example should someone like Dirac or Dicke 
qualify for a higher weighting than a child who dies in infancy before
even learning to count?) I have recently shown~\cite{Carter03} how this 
issue provides insights that are useful for the fundamental problem of the 
interpretation of quantum theory.

\bigskip

\noindent{\bf The strong anthropic principle.}
\medskip
\parindent=2 cm

For the crude qualitative applications of the anthropic principle that 
have been discussed so far in the scientific literature, the fine details 
dealt with by the microanthropic principle~\cite{Carter03} are in practice 
unimportant. 

There is however a refinement of a rather different kind that plays
a significant role in the published literature. This is the distinction 
between what are known as the ``strong'' and ``weak'' versions of the 
anthropic principle. In the ordinary,  widely accepted, ``weak'' version 
the relevant (anthropically weighted) a priori probability is supposed to
concern only a particular given model of the universe, or a part thereof,
with which one may be concerned. In the more controversial ``strong'' 
version the relevant anthropic probability distribution is supposed to 
be extended over an ensemble of cosmological models that are set up with 
a range of different values of what, in a particular model are usually 
postulated to be fundamental constants (such as the well known example of 
the fine structure constant). The observed values of such constants might 
be thereby explicable if it could be shown that other values were 
unfavourable to the existence of anthropic observers. However if (as many 
theoreticians hope) the values of all such constants should turn out to 
be mathematically derivable from some fundamental physical theory, then the 
``strong'' version of the anthropic principle would not be needed.

A prototype example of the application of this ``strong'' kind of
anthropic reasonning was provided by Fred Hoyle's 
observation~\cite{Hoyle53} that the triple alpha process that is
necessary for the formation (from primordial hydrogen and helium) of 
the medium and heavy elements of which we are made is extremely
sensitive to the values of the coupling constants governing the 
relevant thermonuclear reactions in large main sequence stars. This 
contrasts with the case of the biochemical processes (depending notably 
on the special properties of water) to which such considerations do not 
apply, despite the fact that (as discussed by Barrow and 
Tipler~\cite{BarTip86}) they  are also indispensible for our kind of life: 
the relevant biochemical properties are not sensitive to the values of any 
physical coupling parameters but are mathematicaly determined by the 
quantum mechanical consequences of the special properties of the 3 
dimensional rotation group. 

Although it does not affect the chemistry of the light and medium weight
elements that play the dominant role in ordinary biochemistry, the
particular value (approximately 1/137) of the electric coupling parameter
that is (appropriately) known as the ``fine structure'' constant is
more significant for the -- less biologically relevant -- details of heavy 
element chemistry. Of potentially greater ``strong'' anthropic relevance, 
however, is the effect~\cite{Carter74} of the ``fine structure'' constant 
on the convective instabilities that are probably important for the 
creation of planets during main sequence star formation

A particularly topical application~\cite{GarVil03} ~\cite{VilTeg04} of the 
``strong'' anthropic principle concerns the recently estimated value of 
Einstein's cosmological repulsion constant on the supposition that it 
is identifiable with what is commonly referred to as the ``dark energy 
density'' of the universe. If this parameter had been much larger (as might 
have been naively expected from fundamental physical considerations) then 
the universe would have already been inflated to such a low density at such 
an early stage in its life after the big bang that the galactic and stellar 
structures needed for our life systems would never have been able to 
condense out at all.

Although far from tautological, but of considerable scientific interest 
from the point of view of explaining the environment in which we find 
ourselves, the foregoing examples do not actually provide direct 
predictions of facts that are not already well established.  However the 
next section will describe examples in which the anthropic principle 
provides genuine predictions in the form of conclusions that remain 
unconfirmed and even controversial.

\bigskip
\noindent{\bf Anthropic prediction.}
\medskip
\parindent=2 cm

Although oversimplified expressions of the anthropic principle (such as 
the version asserting that life only exists where it can survive)
reduce to mere tautology, the more complete formulation (prescribing
an a priori probability distribution) can provide non-trivial
predictions that may be controversial, and that are subject to rational 
contestation since different from what would be obtained from alternative 
prescriptions for a priori probability, such as the ubiquity principle that 
would attribute a priori (but of course not a posteriori) probability even 
to uninhabited situations.

The example that seems to me most important was  provided by the
prediction~\cite{Carter83} that the occurrence of anthropic observers
would be rare, even on environmentally favorable planets such as ours.
This prediction was based on the observation that our evolutionary 
development on Earth has taken a substantial fraction of the time available 
before our Sun reaches the end of its main sequence (hydrogen burning)
life. This would be inexplicable on the basis of the ubiquity principle, 
which would postulate that the case of our planet was typical and 
hence that life like ours should be common. On the basis of the anthropic 
principle it would also be inexplicable if one supposes that biological 
evolution can procede easily on timescales short compared with those of 
stellar evolution, but it is just what would be expected if the biological 
evolution of life like ours depends on chance events with characteristic 
timescales long compared with those of stellar evolution.

The (as yet unrefuted) implication that I drew from this (more than twenty 
years ago) was that the search for extraterrestrial civilisations was 
unlikely to achieve easy success. I have found however that such 
conclusions tend to be unpopular in many quarters, presumably because they 
involve limitations on the extent and more particularly the duration of 
civilisations such as  ours which (in lieu of personal immortality) many 
people would prefer to think of as everlasting: in the words of Dirac (when 
refusing to accept Dicke's effectively anthropic reasonning~\cite{Dicke61}) 
the assumption to be preferred is ``the one that allows the possibility of 
eternal life''.

One of the most remarkable attempts to show that -- despite the inexorable
action~\cite{Islam77} of the entropy principle commonly known as the 
Second Law of thermodynamics -- life could after all continue to exist 
in the arbitrarily distant future, has been made by Freeman
Dyson~\cite{Dyson79} , whose recent intervention in a related 
debate~\cite{Dyson96} provides  another striking example of the kind of
misunderstanding the anthropic principle was meant to help avoid. However 
the issue on this occasion is not the very long term future of life in the
universe, but the more immediate question of the future of our own 
terrestrial civilisation in the next few centuries. Apparently under the 
influence of wishful thinking reminiscent of Dirac's, Dyson has strongly
objected to a thesis developed particularly by Leslie~\cite{Leslie92}
(and from a slightly different point of view by Gott~\cite{Gott93}) of which 
a conveniently succinct discussion with a comprehensive review of the 
relevant litterature was provided by Demaret and Lambert~\cite{DemLam94}. 
The rather obvious conclusion in question is that the anthropic principle's 
attribution of comparable a priori weighting to comparable individuals 
within our own civilisation makes it unlikely that we are untypical in the 
sense of having been born at an exceptionally early stage in its history, 
and hence unlikely that our civilisation will contain a much larger number 
of people born in the future. 

The foregoing reasonning implies that our numbers will either be cut off
fairly soon by some (presumably ~\cite{Carter83} man made, e.g. ecological) 
catastrophe (the ``doomsday'' scenario~\cite{Leslie92}) or else (more 
``optimistically'') will be subject to a gradual (controlled?) decline that 
must start even sooner but that could be relatively prolonged. Despite the 
fact that such conclusions can be and have been drawn independently (without 
recourse to anthropic reasonning) from other considerations of an economic
or environmental nature, Dyson persists ~\cite{Dyson96} in denying their 
validity, thereby implicitly repudiating the anthropic weighting principle 
on which they are based. Dyson's position seems to be based on what might 
be called the ``autocentric principle'' (the extreme opposite to the 
``ubiquity principle'') as  referred to above, whereby one attributes a 
priori probablity only to one's actual position in the universe.  A 
supposition of this  commonly (but usually subconsciously) adopted kind  
makes it legitimate for Dyson to rule out the use of the Bayes rule as 
something that is redundant (albeit not strictly invalid) because, according 
to this autocentric principle, no a priori probability measure is 
attributable to anything inconsistent with what has already been observed. 
However (quite appart from its failure to face the ecological considerations 
leading to the same conclusions) Dyson's wishful thinking in this context 
seems even less intellectually defensible than Dirac's ubiquitism, because 
the autocentric principle effectively violates Ockham's razor by its 
sollipsistic introduction of an artificial distinction between ``oneself'' 
and other manifestly comparable observers.

\vfill\eject

\end{document}